\documentclass[12pt]{article}
\usepackage[top=25truemm,bottom=25truemm,left=22truemm,right=22truemm]{geometry}

\usepackage{indentfirst}
\usepackage[dvipdfmx]{graphicx}
\usepackage{float}
\usepackage{url}
\usepackage{siunitx}
\usepackage{array}
\usepackage{amsmath}
\usepackage{amssymb}
\usepackage{fancybox}
\usepackage{ascmac}
\usepackage{bm}
\usepackage{mathrsfs}
\usepackage{cancel}
\usepackage{physics}
\usepackage{color}
\usepackage{authblk}
\usepackage{appendix}
\usepackage{comment}
\usepackage{mhchem}
\usepackage{caption,latexsym,amsfonts,mathtools,here}
\usepackage{subcaption}
\usepackage{braket}
\usepackage{hyperref}
\usepackage{multirow}
\usepackage[dvipsnames]{xcolor}
\usepackage{CJKutf8}
\usepackage[
  backend=biber,
  sorting=none,
  style=numeric-comp,
  doi=false, url=true, eprint=true
]{biblatex}
\bibliography{refs}

\numberwithin{equation}{section}

\newcommand{\no}{\nonumber}

\newcommand{\pa}{\partial}

\newcommand{\mpl}{M_{\rm P}}

\begin{document}
\begin{titlepage}
\begin{flushright}
{KUNS-3090}
\end{flushright}

\vspace{50pt}

\begin{center}

{\large{\textbf{Violation of the Leggett-Garg inequality in photon-graviton conversion}}}

\vspace{25pt}
Kimihiro Nomura$^1$, Akira Taniguchi$^2$, and Kazushige Ueda$^{3,4,5}$

\vspace{20pt}

\it {\small 
$^1$Department of Physics, Kyoto University, Kyoto 606-8502, Japan
}\\
\it {\small $^2$Department of Physics, Kyushu University, Fukuoka 819-0395, Japan} \\
\it {\small $^3$National Institute of Technology Tokuyama College, Shunan 745-8585, Japan}\\ 
\it {\small $^4$
Physics Program, Graduate School of Advanced Science and Engineering, Hiroshima University, 1-3-1 Kagamiyama, Higashi-Hiroshima, Hiroshima 739-8526, Japan
}\\
\it {\small $^5$Yasuda Girls' Junior \textsl{\&} Senior High School, Hiroshima 730-0001, Japan}
\end{center}

\vspace{3cm}
\begin{abstract}
The Leggett-Garg inequality (LGI) is a temporal analogue of Bell's inequality and provides a quantitative test of the nonclassicality of a system through its violation. We analytically investigate the violation of the LGI in the context of photon-graviton conversion in a magnetic field background, motivated by its potential applications to testing the nonclassicality of gravity. When gravitational perturbations are quantized as gravitons, the conversion of an initial single photon state gives rise to a superposition of photon and graviton states. We show that the temporal correlations obtained from successive projective measurements on the photon-graviton system violate the LGI. Observation of such a violation would provide a novel avenue for probing the quantum nature of gravity.
\end{abstract}
\end{titlepage}
\setcounter{page}{2}
\tableofcontents
\section{Introduction}
\label{sec:introduction}
Probing the quantumness of gravity remains one of the foremost challenges in modern physics. However, quantum effects in gravity are extremely weak, and the direct detection of a single graviton is considered practically difficult~\cite{Dyson:2013hbl}. This is precisely why more ingenious verification methods are needed to translate the quantum nature of gravity into observable forms.

In recent years, indirect approaches incorporating ideas from quantum optics and quantum information have been actively explored. For gravitational radiation in squeezed states with high occupation numbers, focusing on quantum noise and quantum entanglement induced in detectors offers an indirect means of probing gravitons~\cite{Parikh:2020nrd,Kanno:2020usf,Parikh:2020kfh,Parikh:2020fhy,Kanno:2021gpt,Hsiang:2024qou}. 
Moreover, in the high-frequency regime, where occupation numbers become small, gravitational-wave detection schemes based on graviton-magnon resonance or graviton-photon conversion \cite{Ito:2019wcb,Ito:2022rxn,Ejlli:2019bqj,Berlin:2021txa,Domcke:2022rgu,Kanno:2023whr,Aggarwal:2025noe} may become sensitive to individual gravitons \cite{Carney:2023nzz}. Additional ideas include diagnosing nonclassicality through graviton number statistics exhibiting sub-Poissonian behavior~\cite{Kanno:2018cuk,Giovannini:2019bfw,Giovannini:2019ehc,Kanno:2025fpz}, exploiting nonlinear effects~\cite{Guerreiro:2025sge,Guerreiro:2021qgk,Manikandan:2025dea,Guerreiro:2025mcu,Kanno:2025how,Das:2025kyn}, and employing photon-graviton conversion to search for genuinely quantum graviton states~\cite{Ikeda:2025uae}. The central objective is to test observables that cannot be reproduced by any classical processes.

A useful criterion for experimentally identifying nonclassical behavior in physical systems is provided by Bell's inequality~\cite{Bell:1964kc}, which is based on correlations between spatially separated measurements. Its violation has been confirmed in numerous experiments~\cite{PhysRevLett.49.91,PhysRevLett.49.1804,Tittel:1998ja,Weihs:1998gy,Rowe:2001kop,Salart:2008bxh}, thereby ruling out classical descriptions based on local realism. Bell's inequality has also been considered in cosmology as possible diagnostics of nonclassical correlations in primordial fluctuations~\cite{Ando:2020kdz,Martin:2021znx,Micheli:2022tld,Micheli:2025yux}. Extending the concept of Bell's inequality to temporal correlations leads to the Leggett-Garg inequality (LGI)~\cite{Leggett:1985zz}. The LGI places an upper bound on correlations obtained from measurements performed on a single system at different times. The LGI is derived under two assumptions \cite{Emary:2013wfl}: (i) {\it macroscopic realism} (MR), i.e., the system is always in one of the available states, and (ii) {\it noninvasive measurability} (NIM), i.e., the state can, in principle, be read out without disturbing the subsequent dynamics. 
Both MR and NIM are consistent with our classical intuitions.
Therefore, a violation of the LGI implies that at least one of these assumptions must fail, providing a clear framework for diagnosing nonclassicality. Indeed, LGI tests have been conducted in a wide variety of physical systems, including superconducting qubits~\cite{Palacios-Laloy:2010rdz,Groen:2013uei}, photons~\cite{Goggin:2009rae,Xu:2012ogw,Dressel:2011zz,Suzuki:2012fco,Zhang:2020dva}, and neutrons~\cite{Kreuzgruber:2023hew}. Interestingly, LGI violation due to gravitational interactions has also been discussed~\cite{Matsumura:2021tcc}. More broadly, theoretical studies of LGI violations in harmonic oscillators and quantum fields have been reported~\cite{Hatakeyama:2023kbv,Tani:2023kjh,Hirotani:2024sos}. For a comprehensive review of the LGI, see Ref.~\cite{Emary:2013wfl}.

Based on these considerations, we focus on the LGI as a potential tool for probing the quantum nature of gravity.
Specifically, in this paper, we investigate LGI violation in the context of photon-graviton conversion in a magnetic field background~\cite{gertsenshtein1962wave, Raffelt:1987im}. 
If gravitons exist as quantum excitations of gravitational perturbations, interactions mediated by a magnetic field can generate coherent superpositions of photon and graviton states.
This process can potentially give rise to nonclassical features in temporal correlations, manifested as violations of the LGI.
The LGI is particularly well suited to such two-level systems exhibiting oscillatory dynamics under time evolution and has been widely discussed from a similar perspective in the context of neutrino oscillations~\cite{Gangopadhyay:2013aha, Formaggio:2016cuh,Gangopadhyay:2017nsn,Blasone:2021mbc,Wang:2022tnr}.

This paper is organized as follows.
In Sec.~\ref{sec:LGI}, we briefly review the LGI.
In Sec.~\ref{sec:photon-graviton}, we describe photon-graviton conversion in a constant, homogeneous magnetic field within a quantum-mechanical framework and derive the conversion probability of a photon into a graviton.
In Sec.~\ref{sec:violation}, we show that temporal correlations obtained from successive projective measurements on the photon-graviton system violate the LGI.
Section \ref{sec:conclusion} is devoted to conclusions.
Throughout this paper, we work in natural units, $\hbar = c = 1$.

\section{Leggett-Garg inequality}
\label{sec:LGI}
In this section, we introduce the LGI used in this study.
For detailed reviews, see Ref.~\cite{Emary:2013wfl} and the references therein.

Let us suppose that the system under consideration possesses an observable quantity $Q$ that takes the values $+1$ or $-1$ whenever it is measured.
We denote the result of a measurement at time $t_a$ by $Q_a \equiv Q(t_a)$.
We then define the temporal correlation function $C_{ab}$ between two distinct times $t_a$ and $t_b$ as
\begin{align}
    C_{ab}&\equiv \sum_{Q_a, Q_b=\pm 1} Q_a \, Q_b \, P_{ab}(Q_a, Q_b)\,, 
    \label{eq:correlation}
\end{align}
where $P_{ab}(Q_a, Q_b)$ is the joint probability of obtaining the results $Q_a$ and $Q_b$ from measurements performed at times $t_a$ and $t_b$, respectively.
For instance, experimental runs involving successive measurements at times $t_1$ and $t_2$ (with $t_1 < t_2$) yield the correlation function $C_{12}$. Similarly, runs with successive measurements at $t_2$ and $t_3$ (with $t_2 < t_3$) yield $C_{23}$, while measurements performed at $t_1$ and $t_3$ yield $C_{13}$.
The LGI is derived from two assumptions, macroscopic realism (MR) and noninvasive measurability (NIM), introduced in Sec.~\ref{sec:introduction}.
These assumptions imply that, for any experimental run with a given initial state, a measurement performed at a particular time yields a definite result regardless of whether other measurements are performed before or after that time; that is, each $Q_a$ has a definite value irrespective of which pair $Q_a Q_b$ is measured in the run. 
Consequently, the combination $Q_1 Q_2 + Q_2 Q_3 - Q_1 Q_3$ can take only the values $1$ or $-3$.
Taking the ensemble average over runs, one then obtains the following inequality among the correlation functions, 
\begin{align}
    K_3 \equiv C_{12} + C_{23} - C_{13} \leq 1 \,,
    \label{eq:LGI}
\end{align}
which represents a particular form of the LGI involving measurements at three different times.
(For alternative proofs, see, e.g., Sec.~2.2 of Ref.~\cite{Emary:2013wfl}.)
If one observes $K_3 > 1$, this indicates that at least one of the assumptions of MR and NIM is violated. 
Since both of these assumptions accord with our classical intuition, a violation of Eq.~\eqref{eq:LGI} implies the nonclassicality of the system.
While we use the form of Eq.~\eqref{eq:LGI} throughout this paper, measurements at more than three times can lead to a variety of LGIs \cite{Emary:2013wfl}.

\section{Photon-graviton conversion in quantum mechanics}
\label{sec:photon-graviton}

In this section, we formulate photon-graviton conversion in a constant, homogeneous magnetic field within a quantum-mechanical framework. We first derive the Lagrangian describing photon-graviton conversion, and then perform canonical quantization. This allows us to demonstrate that the time evolution generates superpositions of photon and graviton states. We further show that the corresponding transition probabilities exhibit oscillatory behavior in time.

\subsection{Photon-graviton system}

We start with the Einstein-Hilbert and Maxwell actions,
\begin{eqnarray}
S
=\int d^4 x \, \sqrt{-g}
\left(
\frac{\mpl^2}{2}
\,R
-\frac{1}{4}
F^{\mu\nu}
F_{\mu\nu}
\right)
\label{eq:original action}\,,
\end{eqnarray}
where $\mpl=1/\sqrt{8\pi G}$ denotes the Planck mass,  $R$ is the Ricci scalar associated with the spacetime metric $g_{\mu\nu}$, 
$g$ is the determinant of $g_{\mu\nu}$, and $F_{\mu\nu}$ is the electromagnetic field strength tensor.

We consider electromagnetic waves propagating in a constant, homogeneous background magnetic field.
We let $B^i$ denote the background magnetic field vector, for which the field strength tensor is $\bar{F}_{ij} = \varepsilon_{ijk} B^k$ with $\varepsilon_{ijk}$ being the Levi-Civita tensor.
We also introduce a gauge field $A_\mu$ representing the electromagnetic waves propagating in the background. For the gauge field, we impose the Coulomb gauge condition, ${A^i}_{,i} = A_0 = 0$.
Hereafter, $X_{,i}$ denotes the derivative of $X$ with respect to the spatial coordinate $x^i$.
The gauge field can be expanded in Fourier modes as 
\begin{align}
    A_j(t, \bm{x})=\frac{i}{\sqrt{V}}\sum_{P=+,\times} \sum_{\bm{k}}\, A_{\bm k}^P(t) \, e_j^P(\bm{k}) \, e^{i\bm{k}\cdot \bm{x}}\,,
    \label{eq:Amode}
\end{align}
where the wave vector $\bm{k}$ is discretized as $\bm{k} = 2\pi (n_x, n_y, n_z)/L$, with integers $n_x$, $n_y$, $n_z$, in a box of volume $V = L^3$.
We have introduced the polarization vectors $e_i^P(\bm{k})$ for two polarization modes $P = +, \times$, which satisfy $k^i e_i^P(\bm{k}) = 0$ and 
$e_i^P(\bm{k}) \, e^{Qi}(\bm{k})=\delta^{PQ}$ with the Kronecker delta $\delta^{PQ}$.
We also assume $e_i^+(\bm{k}) = - e_i^+(-\bm{k})$ and $e_i^\times(\bm{k}) = e_i^\times(-\bm{k})$.
Since $A_i(t,\bm{x})$ is a real field, it follows that $A^+_{\bm{k}}(t) = A^{+*}_{-\bm{k}}(t)$ and $A^{\times}_{\bm{k}}(t) = -A^{\times *}_{-\bm{k}}(t)$, where the asterisk (*) denotes complex conjugation.

To study photon-graviton conversion, we further consider gravitational perturbations propagating in Minkowski spacetime. The metric is written as 
\begin{align}
    g_{\mu\nu} dx^\mu dx^\nu =-dt^2+(\delta_{ij}+h_{ij}) dx^idx^j\,.
    \label{eq:metric}
\end{align}
Here, $h_{ij}$ represents the metric perturbation, for which we adopt the transverse-traceless gauge, ${h_{ij}}^{,j}=h^i{}_i=0$. 
Similarly to the gauge field, the metric perturbation can be expanded as
\begin{align}
    h_{jk}(t,\bm{x})=\frac{2}{\mpl}\frac{1}{\sqrt{V}}
    \sum_{P=+,\times}\sum_{\bm{k}} ~h_{\bm{k}}^P(t) \, e_{jk}^P(\bm{k}) \, e^{i\bm{k}\cdot\bm{x}}\, ,
    \label{eq:hmode}
\end{align}
where $e_{ij}^P(\bm{k})$ $(P=+, \times)$ are the polarization tensors defined as $e^+_{ij}(\bm{k}) \equiv [ e^+_i(\bm{k}) \, e^+_j(\bm{k}) - e^\times_i(\bm{k}) \, e^\times_j(\bm{k})] / \sqrt{2}$ and $e^\times_{ij}(\bm{k}) \equiv [ e^+_i(\bm{k}) \, e^\times_j(\bm{k}) + e^\times_i(\bm{k}) \, e^+_j(\bm{k})] / \sqrt{2}$.
The reality of $h_{ij}(t,\bm{x})$ then implies $h^+_{\bm{k}}(t) = h^{+*}_{-\bm{k}}(t)$ and $h^\times_{\bm{k}}(t) = - h^{\times*}_{-\bm{k}}(t)$.

We substitute the decomposition of the electromagnetic field strength, $F_{\mu\nu} = \bar{F}_{\mu\nu} + \partial_\mu A_\nu - \partial_\nu A_\mu$, where  $\bar{F}_{ij} = \varepsilon_{ijk} B^k$ describes the fixed magnetic background and $A_i$ represents the propagating field, together with the metric \eqref{eq:metric}, into the Einstein-Hilbert and Maxwell actions \eqref{eq:original action}. 
We then retain terms up to second order in the dynamical variables $A_i$ and $h_{ij}$, which are expanded as in Eqs.~\eqref{eq:Amode} and \eqref{eq:hmode}.
Without loss of generality, we choose the polarization vector $e^+_i (\bm{k})$ to be orthogonal to the background magnetic field $B^i$.
We then obtain the following quadratic action,
\begin{align}
    S^{(2)}=\frac{1}{2}
    \sum_{P=+,\times}
    \int dt 
    \sum_{\bm{k}} \left[|\dot{h}_{\bm{k}}^P|^2-k^2 |h_{\bm{k}}^P
    |^2+|\dot{A}_{\bm{k}}^P|^2-k^2|A_{\bm{k}}^P|^2 +
    \lambda \,k \, h_{\bm{k}}^P A_{\bm{k}}^{P*}
    + 
    \lambda \, k \, h_{\bm{k}}^{P*} A_{\bm{k}}^P \right]\, ,
    \label{eq:quadraaction}
\end{align}
where an overdot denotes a time derivative, $k \equiv |\bm{k}|$, and $\lambda$ is given by
\begin{align}
    \lambda\equiv\frac{\sqrt{2}}{\mpl}\varepsilon^{ijl} \, e_i^{+}(\bm{k}) \,  \frac{k_j}{k} \, B_l\, ,
    \label{deflambda}
\end{align}
which represents the mixing strength between photons and gravitons induced by the background magnetic field.
To maximize the mixing, we consider electromagnetic waves propagating perpendicular to the background magnetic field. In this case, the mixing strength becomes $\lambda = \sqrt{2} B / \mpl$, where $B \equiv |\bm{B}|$.
Equation \eqref{eq:quadraaction} shows that the two polarization modes $P= +, \times$ are completely decoupled and governed by identical actions. 
To simplify the discussion, hereafter we consider only the $P=+$ polarization mode and omit the polarization index.

Focusing on modes with a given wave vector $\bm{k}$, the dynamics is described by the action $S^{(2)}_{\bm{k}} = \int dt \, L_{\bm{k}}$ with the Lagrangian
\begin{align}
    L_{\bm{k}}= 
    |\dot{h}_{\bm{k}}|^2-k^2|h_{\bm{k}}|^2+|\dot{A}_{\bm{k}}|^2-k^2|A_{\bm{k}}|^2+\lambda \, k \, h_{\bm{k}} A^*_{\bm{k}}+\lambda \, k \, h^*_{\bm{k}}A_{\bm{k}}\, .
    \label{eq:Lbefore}
\end{align}
The last two terms represent the mixing between photons and gravitons and are responsible for photon-graviton conversion.
To solve this system, it is convenient to diagonalize the Lagrangian. Let us introduce new variables $\psi_{+,\bm{k}}$ and $\psi_{-,\bm{k}}$ through the following linear transformation of $A_{\bm{k}}$ and $h_{\bm{k}}$,
\begin{align}
	\begin{pmatrix}
   	\psi_{+,\bm{k}} \\
   	\psi_{-,\bm{k}}
	\end{pmatrix}
	\equiv\frac{1}{\sqrt{2}}\begin{pmatrix}
   	1 & -1 \\
   	1 & 1
	\end{pmatrix}
	\begin{pmatrix}
   	A_{\bm{k}} \\
   	h_{\bm{k}}
	\end{pmatrix}\, .
    \label{eq:eigenstate}
\end{align}
The Lagrangian \eqref{eq:Lbefore} is then rewritten as 
\begin{align}
    L_{\bm{k}}=|\dot{\psi}_{+,\bm{k}}|^2-\Omega_{+,k}^2|\psi_{+,\bm{k}}|^2+|\dot{\psi}_{-,\bm{k}}|^2-\Omega_{-,k}^2|\psi_{-,\bm{k}}|^2\, ,
\end{align}
where $\Omega_{\pm,k} \equiv \sqrt{k^2\pm \lambda \, k}$, with the upper (lower) sign corresponding to the $+$ ($-$) mode.
This Lagrangian is analogous to a system of two decoupled harmonic oscillators, each with angular frequency $\Omega_{\pm,k}$. The conjugate momenta are defined by $\pi_{\pm,\bm{k}}\equiv \pa L_{\bm{k}}/\pa \dot{\psi}_{\pm,\bm{k}}=\dot{\psi}^*_{\pm,\bm{k}}$ and $\pi^*_{\pm,\bm{k}}\equiv \pa L_{\bm{k}}/\pa \dot{\psi}^*_{\pm,\bm{k}}=\dot{\psi}_{\pm,\bm{k}}$. The Legendre transformation gives the Hamiltonian,
\begin{align}
H_{\bm{k}}&=\sum_{s=+,-}\left(\pi_{s,\bm{k}}\dot{\psi}_{s,\bm{k}}+\pi^*_{s,\bm{k}}\dot{\psi}^*_{s,\bm{k}}\right)-L_{\bm{k}}\no\\
&=|\pi_{+,\bm{k}}|^2+\Omega_{+,k}^2|\psi_{+,\bm{k}}|^2+|\pi_{-,\bm{k}}|^2+\Omega_{-,k}^2|\psi_{-,\bm{k}}|^2\, .
\label{eq:Hamilbefore}
\end{align}

Now, we quantize the system in the standard way by promoting the dynamical variables $\psi_{\pm,\bm{k}}$ to operators $\hat{\psi}_{\pm,\bm{k}}$ and expanding them in terms of creation and annihilation operators. Since $\Omega_{\pm,k}$ do not depend on time, the positive and negative frequency solutions are $e^{- i \Omega_{\pm,k}t}$ and $e^{+ i \Omega_{\pm,k}t}$, respectively, and then the operators $\hat{\psi}_{\pm,\bm{k}}(t)$ can be expanded as
\begin{align}
    \hat{\psi}_{\pm,\bm{k}}(t)=\sqrt{\frac{1}{2\Omega_{\pm,k}}}\left(\hat{a}_{\pm,\bm{k}} \, e^{-i\Omega_{\pm,k} t}+\hat{a}^\dagger_{\pm,-\bm{k}} \, e^{+i\Omega_{\pm,k} t}\right)\, .
    \label{eq:modeexpansion}
\end{align}
Here, $\hat{a}_{s,\bm{k}}^\dag$ and $\hat{a}_{s,\bm{k}}$ $(s=+,-)$ are the creation and annihilation operators. We impose the commutation relations $[\hat{a}_{s,\bm{k}}, \hat{a}_{s',\bm{k}'}^\dagger ]=\delta_{s,s'}\delta_{\bm{k},\bm{k}'}$ and $[\hat{a}_{s,\bm{k}}, \hat{a}_{s',\bm{k}'}] = [\hat{a}_{s,\bm{k}}^\dagger, \hat{a}_{s',\bm{k}'}^\dagger]=0$. Substituting the mode expansion~\eqref{eq:modeexpansion} and the corresponding expression for the conjugate momentum $\hat{\pi}_{\pm,\bm{k}}=\dot{\hat{\psi}}_{\pm,\bm{k}}^\dag (= \dot{\hat{\psi}}_{\pm,-\bm{k}})$ into the Hamiltonian~\eqref{eq:Hamilbefore}, we can rewrite $\hat{H}_{\bm{k}}$ in terms of the creation and annihilation operators. 
Although this procedure yields contributions from both the $\bm{k}$ and $-\bm{k}$ modes, we consider only the $\bm{k}$ mode as representative in what follows. Furthermore, a constant term corresponding to the zero-point energy does not contribute to the dynamics and thus can be disregarded. Then, we obtain the Hamiltonian $\hat{H}_{\bm{k}} = \hat{H}_{+,\bm{k}} + \hat{H}_{-,\bm{k}}$ with
\begin{align}
    \hat{H}_{\pm,\bm{k}}=\Omega_{\pm,k} \, \hat{a}_{\pm,\bm{k}}^\dag \hat{a}_{\pm,\bm{k}}\, .
\end{align}
The vacuum state $\ket{0}$ is defined as the state satisfying $\hat{a}_{\pm,\bm{k}} \ket{0} = 0$.
The one-particle states are obtained by acting with the creation operators on the vacuum as $\ket{\psi_{\pm,\bm{k}}}\equiv\hat{a}^\dag_{\pm,\bm{k}}\ket{0}$. Since the one-particle states are eigenstates of $\hat{H}_{\pm,\bm{k}}$ with eigenvalues $\Omega_{\pm,k}$, their time evolution in the Schr\"{o}dinger picture is expressed as
\begin{align}
    \ket{\psi_{\pm,\bm{k}}(t)}=e^{-i\hat{H}_{\pm,\bm{k}}t}\ket{\psi_{\pm,\bm{k}}}=e^{-i\Omega_{\pm,k}t}\ket{\psi_{\pm,\bm{k}}}\, .
    \label{eq:timeevolution}
\end{align}

Note that the photon operator $\hat{A}_{\bm{k}}$ and the graviton operator $\hat{h}_{\bm{k}}$ are related to the operators $\hat{\psi}_{\pm, \bm{k}}$ through the linear transformation \eqref{eq:eigenstate}.
Accordingly, the one-photon state $\ket{A_{\bm{k}}}$ and one-graviton state $\ket{h_{\bm{k}}}$ can be written as superpositions of the states $\ket{\psi_{\pm,\bm{k}}}$ as 
\begin{align}
    \ket{A_{\bm{k}}} &= \frac{1}{\sqrt{2}} ( \ket{\psi_{+,\bm{k}}} + \ket{\psi_{-,\bm{k}}} ) \,, 
    \\
    \ket{h_{\bm{k}}} &= \frac{1}{\sqrt{2}} ( - \ket{\psi_{+,\bm{k}}} + \ket{\psi_{-,\bm{k}}} ) \,.
\end{align}
Since the states $\ket{\psi_{\pm,\bm{k}}}$ acquire different phases during time evolution as shown in Eq.~\eqref{eq:timeevolution}, the superposition of the photon and graviton states evolves nontrivially in time. This leads to photon-graviton conversion, as discussed below.

\subsection{Photon-graviton conversion probability}

We consider an experimental setup in which no graviton excitations are initially present and a photon is incident on a background magnetic field. 
The initial state is thus given by $\ket{\Psi_{\bm{k}}(0)}=\ket{A_{\bm{k}}}=(1/\sqrt{2})\left(\ket{\psi_{+,\bm{k}}}+\ket{\psi_{-,\bm{k}}}\right)$.
From Eq.~\eqref{eq:timeevolution}, this state evolves as $\ket{\Psi_{\bm{k}}(t)} = (1/\sqrt{2}) ( e^{-i\Omega_{+,k} t} \ket{\psi_{+,\bm{k}}} + e^{-i\Omega_{-,k} t} \ket{\psi_{-,\bm{k}}})$.
The probability amplitude for a photon to convert into a graviton at time $t$ is obtained by projecting the time-evolved state $\ket{\Psi_{\bm{k}}(t)}$ onto the graviton state as
    \begin{align}
	\braket{h_{\bm{k}} | \Psi_{\bm{k}}(t)}= i\sin\left(\frac{\Delta \Omega_k}{2} \, t \right) \exp\left(-i \, \frac{\Omega_{+,k}+\Omega_{-,k}}{2} \, t \right)\, .
    \label{eq:convprob1}
	\end{align}
Here, $\Delta \Omega_k\equiv \Omega_{+,k}-\Omega_{-,k}$ denotes the angular frequency splitting.
Since the mixing parameter $\lambda = \sqrt{2} B/ \mpl$ is suppressed by the Planck mass $\mpl$, the condition $\lambda\ll k$ is typically satisfied for realistic magnetic field strengths, implying $\Delta \Omega_k\simeq \lambda$. As a result, the conversion probability from a photon ($\gamma$) to a graviton ($g$) is given by
\begin{align}
    P_{\gamma \to g}(t)&= \left|\braket{h
     | \Psi(t)}\right|^2
    =\sin^2\left(\frac{B}{\sqrt{2}\mpl}t\right) \, ,
    \label{eqPconv}
\end{align}
while the photon survival probability is
\begin{align}
	P_{\gamma \to \gamma}(t)&= \left|\braket{A | \Psi(t)}\right|^2 = \cos^2\left(\frac{B}{\sqrt{2}\mpl}t\right)\, .
    \label{eqPsurv}
\end{align}
These results are consistent with the known expression derived within classical field theory \cite{Masaki:2018eut}.
Here, we have omitted the wave vector label $\bm{k}$, since the results become independent of $\bm{k}$ in the regime $\lambda \ll k$.
These probabilities satisfy $P_{\gamma\to g}(t) + P_{\gamma \to \gamma}(t)=1$ as required by unitarity.
\section{LGI violation in photon-graviton conversion}
\label{sec:violation}

In this section, we examine the violation of the LGI in the context of photon-graviton conversion discussed in the previous section, in order to probe the nonclassicality of the photon-graviton system.

Let the observable $Q(t)$ take the value $Q=+1$ when a photon is observed and $Q=-1$ when a graviton is observed. With this convention, we consider the following temporal correlation function according to Eq.~\eqref{eq:correlation},
\begin{align}
    C_{ab} &= P_{ab}(\gamma, \gamma) - P_{ab}(\gamma, g) - P_{ab}(g, \gamma) + P_{ab}(g, g) \,.
    \label{eq:Cab_def}
\end{align}
Here, for example, $P_{ab}(\gamma, g)$ denotes the joint probability that a photon $(\gamma)$ is observed at time $t_a$ and a graviton $(g)$ is observed at time $t_b$ in a sequence of measurements with $t_a < t_b$.
For the measurement times $t_a$ and $t_b$, we use three combinations, $(a,b) = (1,2), (2,3), (1,3)$.
We consider a sequence of measurements, each of which is projective; that is, if a photon (graviton) is observed at time $t_a$, the state is projected onto the photon state $\ket{A}$ (graviton state $\ket{h}$) at that time.
Then, using the conversion and survival probabilities in Eqs.~\eqref{eqPconv} and \eqref{eqPsurv}, the joint probability $P_{ab}(\gamma, g)$ is given by 
\begin{align}
    P_{ab}(\gamma, g)
    &= P_{\gamma \to \gamma}(t_a) \, P_{\gamma \to g} (t_b - t_a) 
    = \cos^2 \left( \frac{B}{\sqrt{2} \mpl} t_a \right) \sin^2 \left( \frac{B}{\sqrt{2} \mpl} (t_b - t_a) \right) \,.
\end{align}
The remaining joint probabilities $P_{ab}(\gamma, \gamma)$, $P_{ab}(g,\gamma)$, and $P_{ab}(g,g)$ can be obtained in a similar manner.
Consequently, the temporal correlation function \eqref{eq:Cab_def} reduces to
\begin{align}
    C_{ab}&=1-2\sin^2\left(\frac{B}{\sqrt{2}\mpl}(t_b-t_a)\right)\, .
    \label{eq:Cab}
\end{align}
For simplicity, we consider equal temporal separations $t_2-t_1=t_3-t_2=\Delta t$.
In this case, the quantity $K_3$ defined in Eq.~\eqref{eq:LGI} becomes
\begin{align}
    K_3=1-\bigg[4\sin^2\bigg(
    \frac{1}{2}
    \frac{\sqrt{2} \, B}{\mpl}\Delta t\bigg)-2\sin^2\bigg(\frac{\sqrt{2}\,B}{\mpl}\Delta t\bigg)\bigg]\, .
    \label{eq:K3result}
\end{align}
\begin{figure}[t]
    \centering
    \includegraphics[width=9cm]{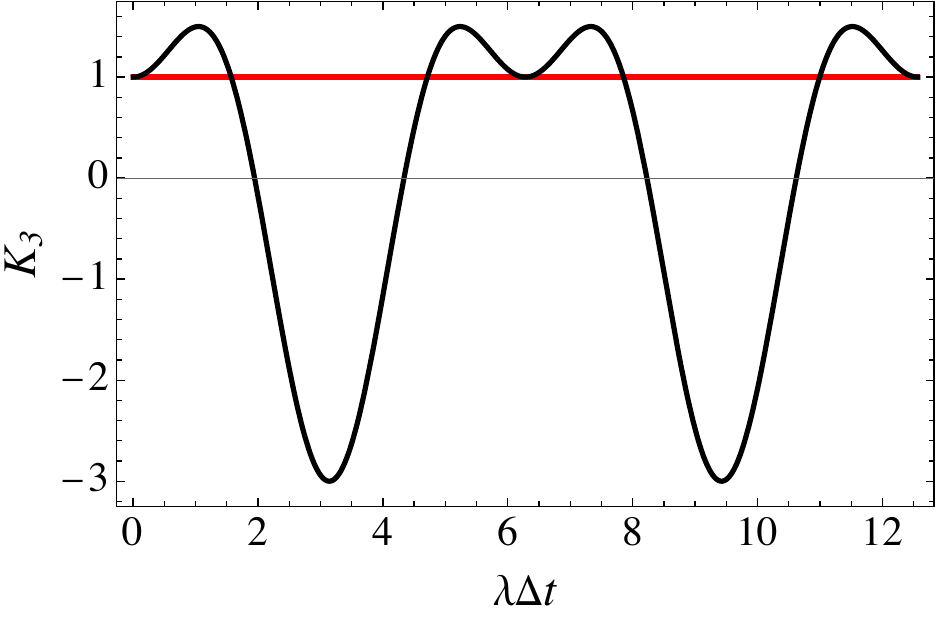}
    \vspace{-2pt}
    \caption{\small The quantity $K_3$ is plotted as a function of the dimensionless parameter $\lambda \,\Delta t$, where $\lambda=\sqrt{2}B/\mpl$.
    The black curve shows the function $K_3$, while the red horizontal line indicates the classical upper bound $K_3=1$ imposed by the LGI. Hence, in the regions where the black curve exceeds the red line ($K_3>1$), the LGI is violated.}
    \vspace{12pt}
    \label{fig:K3}
\end{figure}
The quantity $K_3$ is plotted in Fig.~\ref{fig:K3} as a function of the parameter $\lambda \, \Delta t = (\sqrt{2} B / \mpl) \Delta t$.
As shown in Fig.~\ref{fig:K3}, $K_3$ exhibits oscillatory behavior and exceeds unity ($K_3>1$) over certain ranges of $\lambda \, \Delta t$, implying a violation of the LGI. This indicates that any description compatible with both MR and NIM is ruled out, thereby demonstrating the nonclassicality of photon-graviton conversion.
The maximum value of $K_3$ is 3/2, which is realized at $\lambda \, \Delta t = \pm \pi/3 + 2n\pi$, with integer $n$.
Since photons and gravitons are massless, the temporal separation $\Delta t$ can be replaced by the corresponding propagation length $\Delta L$. Accordingly, the horizontal axis in Fig.~\ref{fig:K3} may also be interpreted as $\lambda \, \Delta L$.

Let us estimate the magnitude of the LGI violation in an experimental setup. 
We consider a strong laboratory-scale magnetic field $B=10~{\rm T}$ and a propagation distance $\Delta L=10~{\rm km}$. 
Expanding Eq.~\eqref{eq:K3result} up to second order in the small parameter $\lambda \, \Delta L$, we obtain
    \begin{align}
    K_3-1\simeq \bigg(\frac{\sqrt{2} \,B}{\mpl} \Delta L\bigg)^2\simeq 3.3 \times 10^{-27}\left(\frac{B}{10\, {\rm T}}\right)^2\left(\frac{\Delta L}{{10\, {\rm km}}}\right)^2\, .
    \label{eqsignalK3}
    \end{align}
This estimate indicates that a sensitivity at the level of $10^{-27}$ would be required to experimentally detect the nonclassicality of photon-graviton conversion in the above setup. 
Moreover, these considerations imply that enhancing the LGI violation requires long-distance propagation in strong magnetic fields.

To relate this estimate to the required magnetic field strength $B$, let us consider the statistical uncertainty in the measured value of $K_3-1$. For temporal correlators $C_{12}$, $C_{23}$, and $C_{13}$ measured using statistically independent data sets with the same number $N$ of runs, the uncertainty can be estimated as (see Appendix)
    \begin{align}
        \Delta (K_3-1)\simeq \sqrt{\frac{6}{N}}\left(\frac{\sqrt{2}B}{\mpl}\Delta L\right)\, .
    \end{align}
Requiring the signal in Eq. \eqref{eqsignalK3} to exceed this statistical uncertainty gives
    \begin{align}
        B\gtrsim 4.3~{\rm T}\left(\frac{10~{\rm km}}{\Delta L}\right)\left(\frac{10^{28}}{N}\right)^{1/2}\, .
    \end{align}
Therefore, for a terrestrial-scale baseline such as $\Delta L\sim 10$ km, a magnetic field of order a few tesla would be sufficient for the signal to overcome the statistical uncertainty, if one could accumulate $N\sim 10^{28}$.

However, accumulating such a large number of measurements would be extremely demanding, and a realistic experiment would also have to control other systematic effects. Thus, while an Earth-based measurement is not excluded in principle, it would be highly challenging in practice and would likely require either enormous photon statistics or some enhancement mechanism, such as using squeezed states~\cite{Ikeda:2025uae}.

\section{Conclusion}
\label{sec:conclusion}
In this study, we have analytically investigated the violation of the Leggett-Garg inequality (LGI) in photon-graviton conversion in a background magnetic field. 
We have shown that, by quantizing gravitational perturbations as gravitons, the time evolution of an initial single-photon state generates a coherent superposition of photon and graviton states.
We have then considered successive projective measurements on the photon-graviton system and evaluated the LGI using the resulting two-time correlation functions. 
We have found violations of the LGI for certain measurement time separations. These violations imply that photon-graviton conversion is incompatible with classical description of gravity that satisfies both macroscopic realism (MR) and noninvasive measurability (NIM). More concretely, an LGI violation can rule out a class of effective descriptions involving a classical gravitational field, in which, for example, the gravitational field induces decoherence of the photon system. In such a description, the density matrix may take the form
    \begin{align}
        \rho(t;h)=p_1(t)\ket{1_\gamma}\bra{1_\gamma}\delta(h-h_1(t))+p_0(t)\ket{0_\gamma}\bra{0_\gamma}\delta(h-h_0(t))\,,
    \end{align}
where $\ket{1_\gamma}$ and $\ket{0_\gamma}$ denote the one-photon and the zero-photon states, respectively. The corresponding classical
gravitational fields are denoted by $h_1(t)$ and $h_0(t)$, respectively. This density matrix represents a mixed state in which the one-photon and zero-photon branches are realized with probabilities $p_1(t)$ and $p_0(t)$, respectively, with $p_1(t)+p_0(t)=1$. 
Then, if we define $Q=+1$ when a photon is present and $Q=-1$ when it is absent, the observable $Q(t)$ can be regarded as having a definite value at any time, regardless of whether a measurement is performed, and an ideal measurement merely reads out this value without disturbing the subsequent evolution. Therefore, this model satisfies both MR and NIM and must satisfy the LGI. The LGI violation predicted in this work indicates that this type of effective classical description is incompatible with quantum photon-graviton conversion.

Our analysis can be further extended by considering a broader range of physical scenarios.
For example, when photons are prepared in a squeezed coherent state and gravitons are in a squeezed state as predicted by primordial gravitational waves, the photon-graviton conversion probability is expected to be enhanced~\cite{Ikeda:2025uae}. 
It is therefore of interest to evaluate LGI violation in such a situation.

Furthermore, to apply our theoretical framework to realistic experimental settings, it would be more promising to reconstruct two-time correlation functions using photon ensembles prepared in the same initial state, rather than performing successive projective measurements on a single photon at different times.
Indeed, in LGI tests based on neutrino oscillations, such an approach has been adopted by assembling measurements from many neutrinos with different energies~\cite{Formaggio:2016cuh,Alam:2026bxn}.
Applying this strategy to photon-graviton conversion would make experimental tests of LGI violations more practically feasible.

\section*{Acknowledgments}
K.N. was supported by JSPS KAKENHI Grant Numbers JP24KJ0117 and JP25K17389.
A.T. was supported by JSPS KAKENHI Grant Number JP25KJ1912.
K.U. was supported by JSPS KAKENHI Grant Number JP24K17050.

\appendix
\section{Statistical error}
In this Appendix, we evaluate the statistical error in the estimate of $K_3$.
Suppose that the temporal correlations $\hat{C}_{ab}$ are estimated from $N_{ab}$ independent runs as
    \begin{align}
        \hat{C}_{ab}\equiv\frac{1}{N_{ab}}\sum_{i=1}^{N_{ab}}Q_a^iQ_b^i\, .
    \end{align}
Since $Q_a^i=\pm 1$ and ($Q_a^i)^2=1$, its variance is
    \begin{align}
        {\rm Var}[\hat{C}_{ab}]=\frac{1-C_{ab}^2}{N_{ab}}\, .
    \end{align}
For the parameter defined as $\hat{K}_3\equiv \hat{C}_{12}+\hat{C}_{23}-\hat{C}_{13}$, its variance is
    \begin{align}
        {\rm Var}[\hat{K}_3]=\frac{1-C_{12}^2}{N_{12}}+\frac{1-C_{23}^2}{N_{23}}+\frac{1-C_{13}^2}{N_{13}}\, .
    \end{align}
Assuming $N_{12}=N_{23}=N_{13}\equiv N$ and $t_2-t_1=t_3-t_2\equiv \Delta L$, we obtain, for $B\Delta L/\mpl\ll 1$,
    \begin{align}
        {\rm Var}[\hat{K}_3]=\frac{1}{N}\left[2\sin^2\left(\frac{\sqrt{2}B}{\mpl}\Delta L\right)+\sin^2\left(\frac{2\sqrt{2}B}{\mpl}\Delta L\right)\right]\simeq \frac{6}{N}\left(\frac{\sqrt{2}B}{\mpl}\Delta L\right)^2\, .
    \end{align}
Thus, the statistical error $\Delta(K_3-1)$ can be estimated as
    \begin{align}
        \Delta (K_3-1)=\sqrt{{\rm Var}[\hat{K}_3]}\simeq \sqrt{\frac{6}{N}}\left(\frac{\sqrt{2}B}{\mpl}\Delta L\right)\, .
    \end{align}

\printbibliography
\end{document}